\documentclass[11pt]{article}

\usepackage{amsfonts}
\usepackage{amsbsy}
\usepackage{epsfig}
\usepackage{latexsym}
\input amssym.def
\input amssym.tex

\def\bequ{\begin{equation}}
\def\eequ{\end{equation}}
\def\barr{\begin{array}}
\def\earr{\end{array}}

\def\ben{\begin{equation}}
\def\een{\end{equation}}
\def\bena{\begin{eqnarray}}
\def\eena{\end{eqnarray}}

\newcommand{\sect}[1]{\setcounter{equation}{0}\section{#1}}

\advance \topmargin by -\headheight
\advance \topmargin by -\headsep
\evensidemargin \oddsidemargin
\advance\hoffset by -3mm  
\advance\voffset by  8mm  
\def\spa#1{\phantom{\fbox{\rule[-#1cm]{0cm}{0cm}}}}

\begin{document}
\hfuzz=100pt
\title{{\Large \bf{Spinning deformations of the D1-D5 system \\ and a geometric resolution of Closed Timelike Curves }}}
\author{\\Carlos A R Herdeiro\footnote{E-mail: crherdei@fc.up.pt}
\\
\\
{{\it Centro de F\'\i sica do Porto}},
\\ {{\it Faculdade de Ci\^encias da Universidade do Porto}},
\\ {{\it Rua do Campo Alegre, 687, 4169-007 Porto, Portugal.}}}

\date{November 30, 2002}
\maketitle


\begin{abstract}
The $SO(4)$ isometry of the extreme Reissner-Nordstrom black hole of ${\mathcal{N}=1}$, $D=5$ supergravity can be partly broken, without breaking any supersymmetry, in two different ways. The ``right'' solution is a rotating black hole (BMPV); the ``left'' is interpreted as a black hole in a G\"odel universe (GBH).  In ten dimensions, both spacetimes are described by deformations of the D1-D5-pp-wave system with the property that the non-trivial Closed Timelike Curves (CTC's) of the five dimensional manifold are \textit{absent} in the universal covering space of the ten dimensional manifold. In the decoupling limit, the BMPV deformation is normalizable. It corresponds to the vev of an IR relevant operator of dimension $\Delta=1$.  The G\"odel deformation is sub-leading in $\alpha'$ unless we take an infinite vorticity limit; in such case it is a non-normalizable perturbation. It corresponds to the insertion of a vector operator of dimension $\Delta=5$. Thus we conclude that from the dual (1+1)-CFT viewpoint the $SO(4)$ R-symmetry is broken `spontaneously' in the BMPV case and  explicitly in the G\"odel case. 
\end{abstract}

\sect{Introduction}
Since the work of Vafa and Strominger \cite{Strominger:1996sh} and Callan and Maldacena \cite{Callan:1996dv}, the D1-D5 system has been a particularly instructive D-brane configuration for the study of black holes in string theory (see \cite{Maldacena:1996ky,Aharony:1999ti,Mandal:2000rp,Peet:2000hn} for reviews). In this paper we give still another example of its richness. We present a simple, supersymmetry preserving, deformation of the D1-D5 system with angular momentum, interpreted in five dimensions as a black hole in Godel's universe (GBH).\footnote{Different supersymmetric deformations of the D1-D5 system with angular momentum have been considered in \cite{Balasubramanian:2000rt,Maldacena:2000dr,Lunin:2001fv}.} The latter is a supersymmetric solution to ${\mathcal{N}=1}$, $D=5$ Supergravity recently found, albeit not interpreted as a black hole in a rotating universe, by Gauntlett et al. \cite{Gauntlett:02}.

G\"odel's universe \cite{godel} has intrigued relativists for over 50 years. It is a solution to Einstein's equations with an apparently harmless type of matter: a pressureless, positive energy density, perfect fluid and a negative cosmological constant. However, the solution is plagued with Closed Timelike Curves (CTC's), which make the Cauchy problem ill-defined and seemingly violate causality. Still, attempts have been made to make sense of spacetimes with CTC's at the classical level \cite{Friedman:xc} and the discussion concerning their consistency shifted to the quantum realm \cite{Hawking:1991nk}.

String theory has already shed some light on this discussion using exactly the D1-D5 system. A five dimensional, supersymmetric, rotating, asymptotically flat black hole solution, dubbed BMPV black hole \cite{BMPV}, has an entropy described in terms of states of the dual (1+1) dimensional CFT, by the AdS/CFT correspondence. As the rotation of the black hole is increased beyond a threshold, we will have `naked' CTC's all over the spacetime; at this point the CFT states one could associate to the spacetime violate the CFT unitarity bound \cite{Herdeiro:00}. Thus, quantum mechanics breaks down when the spacetime becomes a usable `Time machine'.

In an apparently unrelated development \cite{Gauntlett:02}, a G\"odel type solution was found in five dimensions which is {\it{maximally}} supersymmetric. Remarkably, as we shall show, a slight modification of the BMPV black hole describes a black hole in this universe, a solution that preserves one half of the supersymmetry of $D=5$, ${\mathcal{N}}=1$ Supergravity. It follows that the G\"odel universe black hole (GBH) can also be seen as a deformation of the D1-D5-pp-wave system, and that we can use known machinery to analyze this solution from several viewpoints.

We start, in section 2, by reviewing the solutions of Gauntlett et al. \cite{Gauntlett:02} of interest to us. 

In section 3 we will study comparatively the classical five dimensional geometry of the BMPV black hole versus the black hole in G\"odel's universe (GBH).  They belong to the same class of solutions and differ by the choice of a right versus left invariant one-form for squashing a three-sphere. We will discuss the near horizon geometries -two types of `squashed' $AdS_2\times S^3$-, the zero mass limits -G\"odel spacetime and a singular `repulson'- and the supersymmetry of the solutions. Concerning supersymmetry we note that whereas the supersymmetry of the G\"odel black hole is enhanced in the zero mass limit, the one of the BMPV black hole is not. We point out that the entropy of the BMPV black hole depends both on mass and angular momentum, whereas the one of GBH depends solely on mass. From this viewpoint it is reasonable to say that the BMPV is a rotating black hole whereas the GBH is a static black hole in a rotating universe.

In section 4 we uplift the solutions to 10D as deformations of the D1-D5-pp-wave system. We observe that the non-trivial CTC's of the five dimensional solution become topological and hence absent in the universal covering space of the ten dimensional solution. This had already been noticed for the BMPV black hole in \cite{Herdeiro:00} and subsequently arose in the supergravity description of supertubes \cite{Emparan:2001ux}. But it is the first time a geometric `resolution' of CTC's in a G\"odel-type spacetime is given. This resolution is quite a sensitive process. In particular, the non-trivial CTC's are still present in the IIA or M-theory uplifting of the five dimensional G\"odel solution  \cite{Gauntlett:02}. So, the process does not survive T-duality. Moreover, as we shall see, considering a  more general family of five dimensional solutions, the resolution in type IIB does not occur. But it is striking that it works for the two paradigms of spaces with CTC's: a rotating black hole and a G\"odel universe.

By taking the decoupling limit we observe that, in the dual Conformal Field Theory, the BMPV angular momentum is associated to a relevant operator (in the IR), whereas the  G\"odel angular momentum is associated to an  irrelevant one. The latter, however, vanishes as $\alpha'\rightarrow 0$ unless we take a (double) scaling limit of infinite spacetime vorticity. The BMPV spacetime `perturbation' is normalizable and thus it is not a deformation of the (1+1) dimensional CFT, it is a `deformation'  of the states describing the static black hole, which acquire R-charge; thus the SO(4) R-symmetry is broken spontaneously. The G\"odel universe black hole, in the infinite vorticity limit, corresponds to a non-normalizable perturbation, and therefore to a deformation of the (1+1) dimensional CFT by the insertion of the aforementioned irrelevant operator; the SO(4) R-symmetry is explicitly broken.

Five dimensional black holes have been under the spot light recently for a different reason: a counterexample to  black hole uniqueness theorems was found \cite{Emparan:2001wn}. This counterexample pertains uncharged, rotating black holes. Uniqueness seems, however, to cover static \cite{Gibbons:2002av} and supersymmetric \cite{Reall:02} black holes. We believe the supersymmetric black hole in a rotating universe discussed herein is still another example of the surprises and richness of higher dimensional theories. Moreover, it shows how two different black hole spacetimes, both with angular momentum and within the same theory can be very clearly distinguished by the dual string theoretical description.

\sect{Supersymmetric solutions of ${\mathcal{N}}=1$ Supergravity
in D=5}
The minimal supergravity theory in five spacetime dimensions was constructed in \cite{cremmer,Chamseddine:1980sp}. We take the action to be
\bequ
{\mathcal{S}}=\frac{1}{16\pi G_5}\int d^5 x\sqrt{-g}\left(R-F^2-\frac{2}{3\sqrt{3}}\tilde{\epsilon}^{\alpha \beta \gamma \mu \nu}F_{\alpha \beta}F_{\gamma \mu} A_{\nu}\right) \ , \label{5dsugra}\eequ
where $F=dA$, $\tilde{\epsilon}$ is the Levi-Civita tensor, related to the Levi-Civita tensor density by $\tilde{\epsilon}^{\alpha \beta \gamma \delta \mu}=\epsilon^{\alpha \beta \gamma \delta \mu}/\sqrt{-g}$ and we use a `mostly plus' signature. The equations of motion are\bequ R_{\mu \nu}=2\left(F_{\mu \alpha}F_{\nu}^{\ \alpha}-\frac{1}{6}g_{\mu \nu}F^2\right) \ , \ \ \ \ \ D_{\mu}F^{\mu \nu}=\frac{1}{2\sqrt{3}}\tilde{\epsilon}^{\alpha \beta \gamma \mu \nu}F_{\alpha \beta}F_{\gamma \mu} \ . \label{eqmotn1d5}\eequ

 Following Gauntlett et al. \cite{Gauntlett:02}, the
supersymmetric solutions with a timelike Killing vector field of
${\mathcal{N}}=1$ Supergravity in D=5 can be written as\footnote{This solution assumes $f>0$; an analogous set of
solutions exists with $f<0$.} \bequ \barr{l}
\displaystyle{ds^{2}=-f^{2}[dt+\omega]^2+f^{-1}h_{ij}dx^i dx^j,}
\\\\ \displaystyle{F=\frac{\sqrt{3}}{2}d\left(f[dt+\omega]\right)-\frac{1}{\sqrt{3}}G^+ \ . }
\label{gensol}
\earr
\eequ
The five dimensional manifold, ${\mathcal{M}}$, is a non-trivial line bundle,
${\mathcal{M}} \stackrel{\pi}{\longrightarrow} {\mathcal{B}}$, with the following properties:
\begin{description}
\item[a)] The base space, ${\mathcal{B}}$, is a four dimensional hyper-K\"ahler manifold with metric, $h_{ij}$;
\item[b)] f is a globally defined function on ${\mathcal{B}}$ and $\omega$ is a locally defined one-form on ${\mathcal{B}}$. The two form $fd\omega$ is split in its self-dual and anti-self-dual parts (with respect to the hyper-K\"ahler metric) as
\bequ fd\omega=G^- + G^+ \ ; \eequ
\item[c)] The Bianchi identity and equation of motion for the gauge field require \bequ
dG^+=0 \ , \ \ \ \ \ \Delta
f^{-1}=\frac{2}{9}(G^{+})^{ij}(G^{+})_{ij} \ . \label{susyeq} \eequ

\end{description}
Note that if we choose $G^+=0$, that is $d\omega$ is
anti-self-dual, the equations of motion simply require $f^{-1}$ to
be harmonic on ${\mathcal{B}}$. This includes all known
supersymmetric solutions prior to \cite{Gauntlett:02},  with a
timelike Killing vector field, in particular the BMPV black hole \cite{BMPV} (which is a special case of the solutions in \cite{Cvetic:1996xz, Chamseddine:1998yv}).

\subsection{Solutions with flat base space}
We take  ${\mathcal{B}}$ to be $\mathbb{R}^4$, and write its
metric in terms of the left or right invariant one forms (respectively) on
$S^3\simeq SU(2)$: \bequ ds^2_{{\mathbb{R}}^4}
=dr^2+\frac{r^2}{4}\left[(\sigma^1_L)^2+(\sigma^2_L)^2+(\sigma^3_L)^2\right]=dr^2+\frac{r^2}{4}\left[(\sigma^1_R)^2+(\sigma^2_R)^2+(\sigma^3_R)^2\right]
\ . \eequ The explicit form of $\sigma^i_{R,L}$ in terms of Euler
angles can be found in \cite{Gauntlett:02}. Here, we note that
this one-forms obey \bequ
d\sigma^i_R=\frac{1}{2}\epsilon^{ijk}\sigma^j_R\wedge \sigma^k_R \
, \ \ \ \ \ d\sigma^i_L=-\frac{1}{2}\epsilon^{ijk}\sigma^j_L\wedge
\sigma^k_L \ . \eequ The difference in sign is important
in the following. Such difference is associated to the different
sign choice for the two $SU(2)$ algebras in $SO(4)$. In terms of the dual
vector fields to $\sigma^i_{L,R}$, which we denote as
$\xi^{L,R}_i$, we have \bequ \left[\xi^R_i ,
\xi^R_j\right]=-\epsilon_{ijk}\ \xi_k^R \ , \ \ \ \ \left[\xi^L_i
, \xi^L_j\right]=\epsilon_{ijk}\ \xi_k^L \ . \eequ

The natural ansatz for the one-form $\omega$ is 
\bequ
\omega=g(r)\sigma_{L,R} \ , \eequ
where $\sigma_{L,R}$ is any of the left-invariant or any of the right-invariant one-forms; we treat the two cases in parallel. It follows that
\bequ
G^+\equiv \frac{f}{2}(d\omega+\star d\omega)=\frac{f}{2}\left(\dot{g}\pm \frac{2}{r}g\right)\left[dr\wedge \sigma_{L,R}\pm \frac{r}{2}d\sigma_{L,R}\right] \ . \label{G1}\eequ
The top (bottom) signs correspond to right (left) invariant one-forms. This convention will be used throughout the paper. Requiring $dG^+=0$, yields the condition
\bequ
f(r\dot{g}\pm2g)=\ell r^{\pm2} \ , \label{G2}  \eequ
where $\ell$ is a dimensionful constant. Feeding this back in (\ref{G1}), we find, 
\bequ G^+=\pm\frac{\ell}{4}d\left(r^{\pm2}\sigma_{L,R}\right) \ , \ \ \ \Rightarrow \ \ \ (G^{+})^{ij}(G^{+})_{ij}=\frac{4\ell^2r^{\pm4}}{r^4} \ . \label{G3}   \eequ
Solving the remaining equation of motion (\ref{susyeq}) with (\ref{G3}), we obtain \cite{Gauntlett:02}
\bequ
f^{-1}=\left\{\barr{l}\displaystyle{\lambda+\frac{\mu}{r^2}+\frac{\ell^2}{9}r^2} \spa{0.4}\\ \displaystyle{\lambda+\frac{\mu}{r^2}+\frac{\ell^2}{27r^6}} \earr \right. , \ \ \ \ g(r)=\left\{\barr{l}\displaystyle{\frac{j}{r^2}+\ell\left(\frac{\mu}{2}+\frac{\lambda r^2}{4}+\frac{\ell^2r^4}{54}\right) \ \ \ \ \ \ \ \ \ \rm{(R)}} \spa{0.4}\\ \displaystyle{j r^2-\ell\left(\frac{\lambda}{4r^2}+\frac{\mu}{6r^4}+\frac{\ell^2}{270 r^8}\right)\ \ \ \ \ \rm{(L)}}  \earr \right.  \ .  \eequ
$\lambda, \mu, j$ are integration constants. Notice that the dimensions of $j,\ell$ are $[j_R,\ell_L]=L^3$, $[j_L,\ell_R]=L^{-1}$. They are both vorticity (or `angular momentum') parameters since they give non-trivial contributions to the $dtdx^i$ terms in the metric and to the magnetic dipole term $dx^i\wedge dx^j$ in the gauge field. But whereas `$\ell$' contributes as a source for the function $f$, `$j$' decouples from it. 

To write down the two solutions more explicitly we parameterize
$SU(2)$ by Euler angles $(\theta, \phi, \psi)$. The metric on the
base space is then \bequ
ds^2_{{\mathbb{R}}^4}=dr^2+\frac{r^2}{4}\left(d\theta^2+d\psi^2+d\phi^2+2\cos{\theta}d\psi
d\phi\right) \ . \label{R4cartesian} \eequ Pick up the following left and right
invariant one-forms \bequ \sigma^3_L=d\phi+\cos\theta d\psi \ , \ \ \  \sigma^3_R=d\psi+\cos\theta d\phi \ . \eequ  The metric and
gauge potential for the two solutions may then be written 
as 
 \bequ \barr{l}
\displaystyle{ds^{2}=-f^{2}[dt+g\sigma^3_ {L,R}]^2+f^{-1}ds^2_{{\mathbb{R}}^4} \ , \ \ \ \ A=\frac{\sqrt{3}}{2}\left(fdt+\left[fg\mp\frac{\ell}{6}r^{\pm 2}\right]\sigma_{L,R}^3\right) \ . }
\label{solR4}
\earr
\eequ

\sect{Properties of the five dimensional spacetimes}
In the remaining of this paper we will focus on the solutions (\ref{solR4}) with $\ell=0$, since these are sufficiently rich. These solutions are characterized by three parameters: $\mu, j, \lambda$. The right solution is the BMPV black hole. We will interpret the left solution as a black hole in a G\"odel type universe (GBH). For $\lambda=0$ we will be looking at the near horizon geometries of the two types of black holes. For $\mu=0$ we will be looking at their zero mass limits. Let us discuss the several cases in detail.

\subsection{$\lambda=0$; $j,\mu\neq 0$: Squashed $AdS_2\times S^3$ as near horizon geometries}
The solutions reads \bequ \barr{c}
\displaystyle{ds^2=-\frac{r^4}{\mu^2}\left[dt+jr^{\mp2}\sigma^3_{L,R}\right]^2
+\frac{\mu}{r^2}dr^2+\frac{\mu}{4}\left[(\sigma_{L,R}^1)^2+
(\sigma_{L,R}^2)^2+(\sigma_{L,R}^3)^2\right]} \spa{0.4}\\
\displaystyle{A=\frac{\sqrt{3}}{2}\frac{\mu}{r^2}\left[dt+jr^{\mp2}\sigma^3_{L,R}\right]}
\ \earr \ . \label{l0k0} \eequ For $j=0$ this is $AdS_2\times
S^3$ which is the near horizon geometry of the five dimensional
Reissner-Nordstr\"om black hole. Turning on $j$ we deform the
$AdS_2\times S^3$ solutions in two different ways corresponding to
the left and the right solutions. They are still homogeneous
spaces. For the right case this was discussed in \cite{Alonso-Alberca:2002wr}. The isometry group for $AdS_2\times S^3$ includes both
left ($\xi^L_i$) and right ($\xi^R_i$) invariant vector fields on $S^3$.
For $j\neq0$, $\xi^L_1, \xi^L_2$ ($\xi^R_1,\xi^R_2$) are not Killing vector fields
of the left (right) solution (\ref{l0k0}) any longer, since they act as (${\mathcal{L}}$ denotes Lie derivative)
\bequ
{\mathcal{L}}_{\xi^L_i}\sigma^j_ L=\epsilon^{ijk}\sigma^k_ L \ , \ \ \ \ \ {\mathcal{L}}_{\xi^R_i}\sigma^j_ R=-\epsilon^{ijk}\sigma^k_ R \ . \eequ  However the
action of the remaining vector fields is sufficient to ensure that
the isometry group acts transitively on the manifold. In
particular these spacetimes are completely non-singular.

Despite these similarities, the left and right solution are quite
different. Denoting by $\Delta_{L,R}$ the coefficient of
$(\sigma_{L,R}^3)^2$ in (\ref{l0k0}), we have \bequ
\Delta_{L,R}=\frac{\mu}{4}-\frac{j^2}{\mu^2}r^{4\mp4} \ . \eequ
Whereas, for the left solution, this coefficient always becomes negative for sufficiently
large $r$,  in the right solution its ability to change sign depends
crucially in the ratio $4j^2/\mu^3$. The
behavior of the latter is associated to the over-rotating versus
under-rotating regimes of the BMPV black hole (to be reviewed in section 3.3), and one should think about the right solution as the near
horizon geometry of such black hole. The behavior of the left
solution is analogous to the four dimensional G\"odel spacetime.
This is a homogeneous, non-isotropic, rotating, non-expanding
universe. Since $\Delta_L$ becomes negative at large $r$, there
are angular (hence closed) directions becoming timelike, and hence
closed timelike curves (CTC's). Due to the absence of horizons
these can be deformed to pass all over the spacetime leading to
possible causality violations. Note that when $\Delta_{L,R}=0$ the metric is not singular and this is not a null surface; also there is no paralleled propagated singularity when crossing this surface (curvature seen by a freely falling observer), i.e. the components of the Riemann tensor on a parallel propagated frame are well behaved.

It is not a surprise that this G\"odel-type solutions appears as a
deformation of the $AdS_2\times S^3$ vacuum. In fact, the original
G\"odel solution (without the trivial flat direction) can be
thought of as a deformation of $AdS_3$ \cite{Rooman:1998xf}. More specifically, since
$AdS_3\equiv SL(2,{\mathbb{R}})$, we consider the family of
geometries \bequ ds^2=-\alpha^2(\sigma_{L}^0)^2+
(\sigma_{L}^2)^1+(\sigma_{L}^2)^2 \ , \eequ where $\sigma^{\mu}$
are left-invariant one-forms on $SL(2,{\mathbb{R}})$ and $\alpha$
the deformation parameter. For $\alpha^2=1$ we have $AdS_3$. For
$\alpha^2=2$ we have the G\"odel solution. Explicitly this family of metrics can be written as \bequ
ds^2=-\alpha^2\left[dt+\sinh^2\frac{r}{2}
d\phi\right]^2+\frac{1}{4}\left(dr^2+\sinh^2r d\phi^2\right) \ .
\eequ To describe Anti-de-Sitter space the time direction must be periodic,
hence leading to CTC's. However, these are trivial CTC's (topological) which means they are resolved
by going to the universal covering space, $\widetilde{AdS}_3$, for
which $-\infty<t<\infty$. If $\alpha^2>1$,  we
cannot resolve all CTC's even if we go to the universal covering space, since the coefficient of
$d\phi^2$ becomes \bequ \sinh^2r-4\alpha^2\sinh^4{(r/2)}
\ , \eequ therefore negative for sufficiently large $r$. These are
non-trivial CTC's (geometric) since they are homotopic to a point. This is the G\"odel case.

\subsection{$\mu=0$; $\lambda, j\neq 0$:  G\"odel-type universe and Singular Repulson}
The solution is now
\bequ
ds^2=-\left[dt+jr^{\mp2}\sigma^3_{L,R}\right]^2+ds^2_{{\mathbb{R}}^4} \ , \ \ \ \ \ A=\frac{\sqrt{3}}{2}jr^{\mp2}\sigma^3_{L,R} \ . \eequ
Replacing $jr^{\mp2}\rightarrow jr^n$, the Ricci scalar is 
\bequ
R=2j^2(n^2+4)r^{2n-4} \ . \eequ
This is constant everywhere for the left solution. On the other hand, the coefficient of $(\sigma_{L,R}^3)^2$ is 
\bequ
\Delta_{L,R}=\frac{r^2}{4}-\frac{j^2}{r^{\pm4}} \ , \eequ
and hence CTC's will be present at large $r$ for the left solution. Indeed, this is a homogeneous space of the G\"odel type.  Note that for $n \neq 2$ the spacetime is not homogeneous and it is singular at either $r=0$ or infinity. 

The right solution is asymptotically flat an has a timelike curvature singularity at $r=0$. But such singularity is unattainable by freely falling observers. The radial geodesic equation for a point particle of mass $m$, angular momentum $\omega$ (all of it in one 2-plane, the same 2-plane as the background space's $j$), energy E, and affine parameter $\tau$ is
\bequ
\left(\frac{dr}{d\tau}\right)^2=\frac{4E^2}{r^2}\Delta_{L,R}-\left(m^2+\frac{4\omega^2}{r^2}\right)-\frac{8jE\omega}{r^{\pm2+2}} \ . \eequ
For small $r$ the dominating term always becomes negative, hence prohibiting geodesics from entering such region. The geometry is a singular repulson. For the G\"odel case the opposite happens: for large enough $r$ a negative term dominates. In either case a freely falling observer, with $jE\omega<0$ can cross $\Delta_{L,R}$ into the region where CTC's form, but cannot penetrate `too much' into this region. For the physical interpretation and further discussion of the geodesic motion see \cite{gibher}.

\subsection{$j,\mu,\lambda\neq0$: Black Hole in G\"odel universe and BMPV black hole}
Without loss of generality let us take $\lambda=1$, which can
always be achieved by coordinate transformations. It is convenient
to replace the isotropic radial coordinate, $r$, by a
Schwarzschild-type radial coordinate, $\rho$, related by $\rho^2=
r^2+\mu$. The solution is written explicitly as \bequ \barr{c}
\displaystyle{ds^2=-\Delta^2\left[dt+j(\rho^2-\mu)^{\mp1}\sigma^3_{L,R}\right]^2
+\frac{d\rho^2}{\Delta^2}+\frac{\rho^2}{4}\left[(\sigma_{L,R}^1)^2+
(\sigma_{L,R}^2)^2+(\sigma_{L,R}^3)^2\right] } \spa{0.4}\\
\displaystyle{A=\frac{\sqrt{3}}{2}\left[-\frac{\mu}{\rho^2}dt+\frac{j}{\rho^2}(\rho^2-\mu)^{\mp1+1}\sigma_{L,R}^3\right]}
\ . \earr \  \eequ
We have denoted 
\bequ \Delta=1-\frac{\mu}{\rho^2} \ . \eequ For $j=0$ this is the five dimensional extreme
Reissner-Nordstr\"om solution. There is a timelike curvature
singularity at $\rho=0$ and a degenerate black hole event horizon at
$\rho_H=\sqrt{\mu}$. The spacetime is asymptotically flat with the
typical Carter-Penrose diagram for extreme RN black holes. For
$j\neq0$ we still have a curvature singularity at $\rho=0$, since the Ricci scalar is
\bequ
R=\frac{16j^2(\rho^2-\mu)^{2\mp2}-2\mu^2\rho^2}{\rho^8} \ . \eequ
The coefficient of $(\sigma^3_{L,R})^2$ in the metric
becomes \bequ \Delta_{L,R}(\rho)=\left\{\barr{c}
\displaystyle{\ \ \ \ \ \ \frac{\rho^2}{4}-\frac{j^2}{\rho^4} \ \ \ \ \ \ \ \ \ (R)}\spa{0.4}\\
\displaystyle{\frac{\rho^2}{4}-j^2\left(\rho-\frac{\mu}{\rho}\right)^4 \ \
(L) } \earr \right. \ . \label{deltalr} \eequ The solution is invariant under
$j,t\rightarrow -j,-t$. So we take $j$ positive. Define $\rho_{C}$ by $\Delta_{L,R}(\rho_{C})=0$, which is the boundary of the region where angular directions become timelike.  

The right
solution is the BMPV spacetime: a rotating, asymptotically flat black hole. We have CTC's in the region $\rho^3<2j\equiv \rho_{C}^3$. If $\rho_{C}<\rho_H$ the CTC's are hidden behind the horizon. This is the under-rotating regime. For $\rho_{C}>\rho_H$, the CTC's are naked, and can be deformed to pass through any spacetime point. The surface $\rho=\rho_H$ is now timelike; it is highly repulsive, and no causal geodesic can go beyond it, becoming an effective boundary for the spacetime \cite{gibher}. This is the over-rotating regime, which is similar to the repulson described in the last subsection with the crucial difference that the repulsive surface is now $\rho=\rho_H\neq 0$, hence non-singular.  Since the spacetime is asymptotically flat one can define ADM mass and angular momentum vector (corresponding to the two independent two-planes) of the BMPV black hole. One obtains \cite{GMTown}
\bequ
M=\frac{3\pi \mu}{4G_5} \ , \ \  \ \  \vec{J}=\left(0,\frac{j\pi}{2G_5}\right) \ . \label{mjbmpv} \eequ The entropy of this black hole is 
\bequ
S_{BMPV}=\frac{\pi^2}{2G_5}\sqrt{\mu^3-4j^2} \ , \eequ
which becomes ill defined in the over-rotating case.  A very curious property of the BMPV spacetime is that the angular velocity of the horizon is zero \cite{GMTown}. This is a necessary condition for supersymmetry, since susy is incompatible with an ergo-region.

The left solution has CTC's both for large and for small radial coordinate, since the second term in (\ref{deltalr}) dominates in both regimes. But $\Delta_L$ is not always negative. At $\rho=\sqrt{\mu}$ it is positive. Can this surface be interpreted as a horizon? It is certainly a null surface, as can be seen by a standard analysis. First introduce regular coordinates on the horizon. This is achieved by choosing a retarded time and a new angular coordinate,
\bequ
dv=dt+a(\rho)d\rho \ , \ \ \ \ \left\{\barr{l}d\tilde{\psi}=d\psi+b(\rho)d\rho  \spa{0.3}\\ d\tilde{\phi}=d\phi+b(\rho)d\rho  \earr \right.  \ \Rightarrow \ \ \ \tilde{\sigma}^3_{L,R}=\left\{\barr{c} d\tilde{\psi}+\cos{\theta}d\phi \spa{0.3}\\ 
d\tilde{\phi}+\cos{\theta}d\psi \earr \right. =\sigma^3_{L,R}+b(\rho)d\rho \ . \eequ
(Top/bottom refer to Right/Left solutions). The functions $a(\rho), b(\rho)$ are chosen as to eliminate the $d\rho^2$ and $d\rho \tilde{\sigma}^3_{L,R}$ terms in the new metric. This requirement yields
\bequ
a(\rho)=\frac{2\rho^3\sqrt{\Delta_{L,R}}}{(\rho^2-\mu)^2} \ , \ \ \ \ \ b(\rho)=\frac{2j(\rho^2-\mu)^{\mp1}}{\rho\sqrt{\Delta_{L,R}}}  \ . \eequ
The metric becomes
\bequ ds^2=dv\left[-\Delta^2 dv+\frac{\rho}{\sqrt{\Delta_{L,R}}}d\rho-2j\Delta^{2\mp1}\rho^{\mp2}\tilde{\sigma}^3_{L,R}\right]+\frac{\rho^2}{4}\left[(\sigma_{L,R}^1)^2+
(\sigma_{L,R}^2)^2\right]+\Delta_{L,R}(\tilde{\sigma}_{L,R}^3)^2 \ . \eequ We want the new coordinates to cover the surface $\rho=\sqrt{\mu}$. This requires $\Delta_{L,R}(\sqrt{\mu})>0$. Whereas for the right solution this is only true for the under-rotating case, for the left solution this is true for any value of $j$. Consider now the family of surfaces $S(\rho)=0$. The normal vector to the surface has norm 
\bequ
\partial_{\mu}S\partial_{\nu}S g^{\mu \nu}=g^{\rho \rho}= \Delta^2 \ . \eequ The surface at $\rho=\sqrt{\mu}$ is therefore null. Hence the solution is rightly interpreted as a black hole in G\"odel's Universe. Its entropy, computed from the horizon area, is the same as that of a static black hole
\bequ
S_{GodelBH}=\frac{\pi^2}{2G_5}\sqrt{\mu^3} \ . \eequ
Both entropies will be analyzed from the viewpoint of the D1-D5 system in section 4.

\subsection{Supersymmetry}
The solution (\ref{solR4}) with $\ell=0$ takes the form
\bequ ds^2=-({\bf e}^0)^2+\delta_{ij}{\bf e}^i {\bf e}^j \ , \ \ \ \ F=\frac{\sqrt{3}}{2}d{\bf e}^0 \ , \eequ with the frames 
\bequ 
{\bf e}^0=f(dt+\omega) \ , \ \ \ \ \ {\bf e}^i=f^{-1/2}dx^i \ . \eequ
The Killing spinor equation for (\ref{5dsugra}) yields \cite{Herdeiro:00}
\bequ
\barr{l}
\displaystyle{0=d\epsilon +\frac{1}{4}{\bf \omega}_{ab}\Gamma^{ab}\epsilon +\frac{i}{4\sqrt{3}}\left({\bf e}^a\Gamma^{bc}_{\ \ \ a}-4{\bf e}^b\Gamma^c\right)F_{bc}\epsilon}
\\\\\ \ \ \ \  = \displaystyle{{\bf e}^0\left(\frac{\partial_t \epsilon}{f} +\left[\frac{f^2 a_{ij}\Gamma^{ij}}{8}+\frac{i\partial_{i}f \Gamma^{i}}{2f^{1/2}}\right]\left(1-i\Gamma^{0}\right)\epsilon\right)+{\bf e}^k\left(f^{1/2}\left(\partial_k-\omega_k \partial_t\right)\epsilon -\frac{\partial_k f}{2f^{1/2}}i\Gamma^{0}\epsilon + \right.}
\\\\ \ \ \ \ \ \ \ \ + \displaystyle{\left. \left[\frac{\partial_i f \Gamma^{i}_{ \ k}}{4f^{1/2}}+\frac{f^2 a_{ki}\Gamma^{0i}}{2}\right]\left(1-i\Gamma^{0}\right)\epsilon -\frac{f^2\Gamma^{0i}}{4}\left(a_{ki}+\star a_{ki}\right)\epsilon\right).}
\earr
\eequ
We have denoted by $a_{ij}$ the components of the two form $a=d\omega$ and by $\star a$ the Hodge dual of $a$ on ${\Bbb R}^4$.\footnote{We have taken $\Gamma_{01234}=i$.} If we take $a$ to be anti-self-dual on  ${\Bbb R}^4$ the last term vanishes and we find the obvious set of Killing spinors
\bequ
\epsilon =f^{\frac{1}{2}}\epsilon_0, \ \ \  \epsilon_0=i\Gamma^0 \epsilon_0 \ , \label{kspinors1}
\eequ
corresponding to four independent Killing spinors (one half of the vacuum supersymmetry). This is the supersymmetry preserved by the solutions in section 3.3: the black hole in G\"odel's universe and the BMPV black hole.

Under which conditions can we have extra Killing spinors? Take the remaining set of spinors 
\bequ
\epsilon=\chi(x^{\mu})\epsilon_0 \ , \ \ \ \ \epsilon_0=-i\Gamma^0 \epsilon_0 \ , \eequ
where $\chi(x^{\mu})$ is some matrix, to find the condition
\bequ
\barr{l}
\displaystyle{0={\bf e}^0\left(\frac{\partial_t \chi}{f} +\frac{i\chi \partial_{i}f \Gamma^{i}}{f^{1/2}}\right)\epsilon_0+{\bf e}^k\left(f^{1/2}\left(\partial_k-\omega_k \partial_t\right)\chi +\frac{\partial_k f}{2f^{1/2}}\chi + \left[\frac{\partial_i f \Gamma^{i}_{ \ k}}{2f^{1/2}}-if^2 a_{ki}\Gamma^{i}\right]\chi \right)\epsilon_0 \ .}
\earr \label{extraks}
\eequ

If we do not allow the spinor to be time dependent, $\chi=\chi(x^i)$, we find that $f$ must be a constant, which we take to be unity, and the condition \bequ
\partial_k\chi=ia_{ki}\Gamma^i \chi \label{condition1} \ . 
\eequ
A constant $f$ corresponds to the solutions with $\mu=0$ in section 3.2, thus we now ask if we can solve (\ref{condition1}) for such solutions. First notice that cartesian coordinates on ${\mathbb{R}^4}$ are related to the ones of the form (\ref{R4cartesian}) by
\bequ
x^1+ix^2=r\cos{\left(\theta/2\right)}e^{i(\phi+\psi)/2} \ , \ \ \ x^3+ix^4=r\sin{\left(\theta/2\right)}e^{i(\psi-\phi)/2}\ . \eequ
Thus, in cartesian coordinates 
\bequ \sigma^3_{L,R}=\frac{2}{r^2}[x^1dx^2-x^2dx^1\pm(x^3dx^4-x^4dx^3)]\equiv \frac{2}{r^2}(I^3_{L,R})_{ij}x^idx^j \ , \eequ
which defines the constant antisymmetric matrices $(I^3_{L,R})_{ij}$ and \bequ
\omega=2jr^{\mp2-2}(I^3_{L,R})_{ij}x^idx^j \ \Rightarrow \ \ a=d\omega=\left\{ \barr{c} \displaystyle{\frac{2j}{r^{4}}(I^3_R)_{ij}\left\{\delta^i_k-4\frac{x_kx^i}{r^2}\right\}dx^k\wedge dx^j} \spa{0.5}\\ \displaystyle{2j(I^3_L)_{ij}dx^i\wedge dx^j} \earr \right. \ . \eequ
It is straightforward to check that $a=-\star a$ for both cases. 

For the left case $a$ is a constant matrix and (\ref{condition1}) becomes
\bequ
\partial_k\chi=i4j(I^3_ L)_{ki}\Gamma^i\chi \ \ \Rightarrow \ \ \chi=\exp{(i4j(I^3_ L)_{ki}\Gamma^ix^k)} \ . \eequ
This shows the G\"odel type universe of section 3.2 is maximally supersymmetric \cite{Gauntlett:02}.

For the right case one can check that the integrability condition for (\ref{condition1}) 
\bequ
\partial_{[l}a_{k]j}\Gamma^j \chi=0  \ , \eequ
is not obeyed. Hence, the singular repulson of section 2.3 preserves only half of the vacuum supersymmetry.

Although we have not checked, it is natural that supersymmetry is also going to be enhanced in the near horizon geometry for both solutions. For the right solution this has been checked in \cite{Kallosh:1996vy, GMTown}. For the left solution this would yield another maximally supersymmetric solution of the G\"odel type, (\ref{l0k0}).

\sect{Ten dimensional interpretation in IIB String Theory}
Consider type IIB supergravity with only the graviton, dilaton and Ramond-Ramond two form potential being excited. The equations of motion are
\bequ
\barr{c} \displaystyle{D_M\partial^M \phi =\frac{e^{\phi}}{12}F_{MNP}F^{FNP} \ ,\ \ \ \ \ D_M(e^{\phi} F^{MNP})=0 \ ,} \spa{0.4}\\
\displaystyle{R_{MN}=\frac{1}{2}\partial_{M}\phi \partial_{N}\phi+\frac{e^{\phi}}{4}F_{MPQ}F_{N}^{\ PQ}-\frac{1}{48}g_{MN}e^{\phi}F_{PQR}F^{PQR} \ ,} \earr  \label{eqmotiib}\eequ
which can be derived from the action (which is therefore a consistent truncation of IIB Sugra)
\bequ
\mathcal{S}^{(10)}=\frac{1}{2\kappa^2}\int d^{10}x\sqrt{-g}\left(R-\frac{1}{2}\partial_{M}\phi\partial^{M}\phi-\frac{1}{2\cdot 3!}e^{\phi}F_{MNP}F^{MNP}\right) \ .\eequ
As usual $F=dC$. Split the ten-dimensional coordinates as $x^M=(x^{\mu},y^i)$, where $\mu=0 \dots 4$, $i=1\dots 5$. The vectors $\partial/\partial y^i$ are assumed to be Killing vector fields. Then, perform a Kaluza-Klein reduction with the ansatz,
\bequ \barr{c} \displaystyle{
ds^2=e^{a\hat{\lambda}(x)}\left[e^{b\hat{\psi}(x)}\hat{g}_{\mu \nu}dx^{\mu} dx^{\nu}+e^{-4a\hat{\lambda}(x)}\left(dy^1+\hat{A}^{(1)}_{\mu}dx^{\mu}\right)^2\right]+e^{-3b\hat{\psi}(x)/4}ds^2({\mathbb{T}^4}) \ ,} \spa{0.4}\\ \displaystyle{C=\frac{1}{2}\hat{B}_{\mu\nu}dx^{\mu}\wedge dx^{\nu}+\hat{A}^{(2)}_{\mu}dx^{\mu}\wedge dy^1 \ , \ \ \ \ \ \ \phi=\hat{\phi}(x) \ .} \earr \eequ
Hatted quantities are five dimensional, $ds^2({\mathbb{T}^4})$ is the metric on a flat four-torus and we have singled out one of the extra-dimensions, $y^1$, which can have off diagonal terms in the metric and non trivial legs in the Ramond-Ramond field, hence giving rise to two extra gauge fields. The powers of the several exponential factors in the metric ansatz are chosen as to obtain the five dimensional Einstein frame, keeping two constants, $a,b$, which can be chosen as to yield canonical normalization for the kinetic terms of the scalar fields. Such choice gives $117 b^2=8$, $6a^2=1$. It is also convenient to define the new scalar $\chi(x)=b\psi(x)/a-\lambda(x)$. We obtain the following five dimensional action
\bequ \barr{c}
\displaystyle{\mathcal{S}^{(5)}=\frac{1}{2\hat{\kappa}^2}\int d^{5}x\sqrt{-\hat{g}}\left(\hat{R}-\frac{1}{2}\partial_{\mu}\hat{\phi}\partial^{\mu}\hat{\phi}-\frac{1}{2}\partial_{\mu}\hat{\chi}\partial^{\mu}\hat{\chi}-\frac{1}{2}\partial_{\mu}\hat{\psi}\partial^{\mu}\hat{\psi}-\frac{1}{2\cdot 3!}e^{\hat{\phi}+2a\hat{\chi}-4b\hat{\psi}}\hat{H}_{\mu \nu \alpha}\hat{H}^{\mu \nu \alpha} \right.} \spa{0.4}\\
\displaystyle{\ \ \ \ \ \ \ \left.-\frac{1}{4}e^{4a\hat{\chi}-5b\hat{\psi}}\hat{F}^{(1)}_{\mu \nu}\hat{F}^{(1)\mu \nu}-\frac{1}{4}e^{\hat{\phi}-2a\hat{\chi}+b\hat{\psi}}\hat{F}^{(2)}_{\mu \nu}\hat{F}^{(2)\mu \nu}\right) \ ,} \earr \ \label{action5d} \eequ
The field strengths are defined as $\hat{F}^{(i)}=\hat{A}^{(i)}$ and $\hat{H}=d\hat{B}-\hat{A}^{(1)}\wedge d\hat{A}^{(2)}$, where the latter combination naturally arises in the reduction procedure. Moreover, denoting the volume of ${\mathbb{T}^4}$ by $(2\pi)^4V$ and the radius of the remaining circle of compactification by $R$ we have $(2\pi)^5VR\hat{\kappa}^2=\kappa^2$. 

Requiring all scalars to be constant and $\hat{A}^{(1)}=\hat{A}^{(2)}\equiv \hat{A}$ (denote $\hat{F}=d\hat{A}$), the equations of motion of (\ref{action5d}) yield the following conditions
\bequ \barr{c}
\displaystyle{\hat{H}_{\mu \nu \alpha}\hat{H}^{\mu \nu \alpha}=-3\hat{F}_{\mu \nu}\hat{F}^{\mu \nu} \ , \ \ \ D_{\alpha}\hat{F}^{\alpha \beta}=-\frac{1}{2}\hat{H}^{\beta \mu \nu}\hat{F}_{\mu \nu} \ ,\ \ \ D_{\mu}\hat{H}^{\mu \alpha \beta}=0 \ , } \spa{0.4}\\ \displaystyle{\hat{G}_{\mu \nu}=\hat{F}_{\mu \alpha}F_{\nu}^{\ \alpha}-\frac{1}{4}\hat{g}_{\mu \nu}\hat{F}_{\alpha \beta}\hat{F}^{\alpha \beta}+\frac{1}{4}\left(\hat{H}_{\mu \alpha \beta}\hat{H}_{\nu}^{\ \alpha \beta}-\frac{1}{6}\hat{g}_{\mu \nu}\hat{H}_{ \alpha \beta \gamma}\hat{H}^{\alpha \beta \gamma}\right) \ . } \earr \eequ
Further imposing \bequ
\hat{H}=-\hat{\star} d\hat{A} \ , \ \ \ \ \Rightarrow \ \ \ \ d\hat{B}=\hat{A}\wedge d\hat{A} -\hat{\star} d\hat{A} \ , \eequ and performing the rescaling $\hat{A}\rightarrow 2\hat{A}/\sqrt{3}$ we recover the equations of motion of ${\mathcal{N}}=1$, $D=5$ supergravity, (\ref{eqmotn1d5}).`$\hat{\star}$' denotes Hodge duality with respect to $\hat{g}$. Hence, a solution, $(\hat{g},\hat{A})$ of the five dimensional supergravity theory uplifts as the solution
\bequ \barr{c} \displaystyle{
  ds^2=\hat{g}_{\mu \nu}dx^{\mu}dx^{\nu}+\left(dy^1+\frac{2}{\sqrt{3}}\hat{A}_{\mu}dx^{\mu}\right)^2+ds^2({\mathbb{T}}^4) \ , } \spa{0.4}\\ \displaystyle{H=dC=\frac{2}{\sqrt{3}}d\hat{A}\wedge \left(dy^1+\frac{2}{\sqrt{3}}\hat{A}\right)-\frac{2}{\sqrt{3}}\hat{\star}d\hat{A} \ , } \earr \eequ
of type IIB (or IIA or I, since the dilaton decouples) supergravity. We apply this result to (\ref{gensol}), taking the gauge potential to be 
\bequ
\hat{A}=\frac{\sqrt{3}}{2}\left((f-1)dt+f\omega-\frac{2}{3}h^+\right) \ , \eequ
where $h^+$ is the potential for the closed form $G^+$, $dh^+=G^+$. We obtain the ten dimensional solution
\bequ \barr{c} \displaystyle{ds^2=f\left[-dt^2+(dy^1)^2-\left(1-f^{-1}\right)(dt-dy^1)^2+2\omega\left(dy^1-dt-\frac{2}{3}h^+\right) \right.} \spa{0.3}\\ \displaystyle{ \left. ~~~~~~~~~~~~~~ -\frac{4}{3}h^+f^{-1}\left(dy^1-(1-f)dt-\frac{1}{3}h^+\right) \right]+f^{-1}h_{ij}dx^idx^j+ds^2({\mathbb{T}^4})} \ , \spa{0.4}\\ \displaystyle{H=d\left[f(dt+\omega)-\frac{2}{3}h^+\right]\wedge \left(dy^1-dt+f(dt+\omega)-\frac{2}{3}h^+\right)-\hat{\star}d\left[f(dt+\omega)-\frac{2}{3}h^+\right]} \ . \label{10dsol} \earr  \eequ

\subsection{Higher dimensional interpretation of BMPV and G\"odel black hole}
Start by specializing (\ref{10dsol}) to (\ref{solR4}) with $\ell=0$. We find
\bequ \barr{c} \displaystyle{ds^2=f\left[-dt^2+(dy^1)^2-\left(1-f^{-1}\right)(dt-dy^1)^2+\frac{2j}{r^{\pm2}}(dy^1-dt)\sigma_{L,R}^3  \right]+f^{-1}ds^2_{\mathbb{R}^4}+ds^2({\mathbb{T}^4})} \spa{0.6}\\ \displaystyle{C=f\left(dt+\frac{j}{r^{\pm2}}\sigma_{L,R}^3\right)\wedge \left(dy^1-dt\right)+\frac{\mu}{4}\cos{\theta}d\phi\wedge d\psi} \earr \ . \label{10dsolequal} \eequ
The right solution is a special case of a more general solution with three different charges found in \cite{Herdeiro:00}. This suggests  the following solution
of type IIB supergravity
\bequ
\barr{c}
\displaystyle{ds^2_E=f_5^{-\frac{1}{4}}f_1^{-\frac{3}{4}}\left[-dt^2+(dy^1)^2+f_K(dt-dy^1)^2 +\frac{2j}{r^{\pm2}}(dy^1-dt)\sigma^3_{L,R}   \right]+f_5^{\frac{3}{4}}f_1^{\frac{1}{4}} ds^2_{\mathbb{R}^4}+\left(\frac{f_1}{f_5}\right)^{\frac{1}{4}} ds^2({\mathbb{T}^4})}\ ,
\spa{0.6}\\
\displaystyle{C_{RR}=f_1^{-1}\left(dt+\frac{j}{r^{\pm2}}\sigma^3_{L,R}\right)\wedge (dy^1-dt)+\frac{P}{4}\cos{\theta} d\phi\wedge d\psi \ , \ \ \ \ \ e^{-2(\phi-\phi_\infty)}=\frac{f_5}{f_1}.}
\label{d15bri}
\earr
\eequ
The three functions $f_1$, $f_5$, $f_K$ are given by:
\bequ
f_5=1+\frac{P}{r^2}, \ \ \ \ \ \ \  \ \ f_1=1+\frac{Q}{r^2}, \ \ \ \ \ \ \ \ \ f_K=\frac{Q_{KK}}{r^2}.
\eequ
It is straightforward to verify this indeed solves (\ref{eqmotiib}) for both left and right cases. Setting $j=0$ we recover the standard D1-D5-pp wave system \cite{Callan:1996dv}. Setting $P=Q_{KK}=Q\equiv \mu$, and denoting $f_1=f_5=1+f_{K}\equiv f^{-1}$ we recover (\ref{10dsolequal}).

For the `right' solution (\ref{d15bri}) can be interpreted as a $D1$-brane inside a $D5$ with a Brinkmann wave propagating along the string. For the `left' solution one should think of (\ref{d15bri}) as the D1-D5-pp-wave system in a rotating background, which is not asymptotically flat. 

\subsection{A resolution of causality violations}
It was shown in \cite{Herdeiro:00}  that the non-trivial CTC's of the BMPV black hole become trivial in its ten dimensional description, by virtue of the Kaluza-Klein `oxidation'. This is also true for the supersymmetric G\"odel solution. Examining (\ref{10dsolequal}) there are no obvious CTC's; one does not  identify a periodic direction that becomes timelike. However, the vector
\bequ
{\bf k}=\alpha \partial/\partial y^1 +\beta \xi^{L,R}_3 \ , \eequ
which is a linear combination of two spacelike vectors, has norm
\bequ |{\bf k}|^2=\frac{r^2}{4f}\alpha^2+\frac{2jf}{r^{\pm2}}\alpha \beta +\beta^2 \ . \eequ
For $j=0$ this is non-negative for any choice of $\alpha,\beta$, at any spacetime point. For $j\neq 0$, asking when ${\bf k}$ becomes null yields a quadratic equation in $\beta/\alpha$ with discriminant binomial $-\Delta_{L,R}$ (given by (\ref{deltalr})). Hence, there will be real solutions for $\Delta_{L,R}\le 0$;  for $\Delta_{L,R}<0$ we can have a timelike ${\bf k}$. Thus we can find timelike curves, with tangent vector ${\bf k}$,  when $\Delta_{L,R}<0$. However these curves can not be closed until we make the $y^1$ direction compact. These are the CTC's seen in five dimensions, with the crucial difference that they are not homotopic to a point any longer; they are trivial, since they are resolved in the universal covering space of the manifold. In \cite{Herdeiro:00} a pictorial description of the resolution process is given.

If we apply T-duality along the $y^1$ direction to (\ref{d15bri}) we obtain the IIA (string frame) geometry
\bequ
ds^2=-\frac{(f_1f_5)^{-1/2}}{1+f_K}\left[dt+\frac{j}{r^{\pm2}}\sigma_{L,R}^3\right]^2+(f_1f_5)^{1/2}\left( \frac{(dy^1)^2}{1+f_K}+ds^2_{{\mathbb{R}}^4}\right)+\left(\frac{f_1}{f_5}\right)^{1/2}ds^2({\mathbb{T}^4}) \ , \label{D0D4} \eequ
which has non-trivial CTC's. The T-duality has exactly canceled the effect of the Kaluza-Klein `oxidation' and that is the reason why the M-theory uplifting of the G\"odel solution studied in \cite{Gauntlett:02} still exhibits the same type of Closed Timelike Curves as in five dimensions.

Solution (\ref{D0D4}) is a G\"odel type universe in the limit $f_1=f_5=1$, $f_k=0$, and it is a delocalized (in the $y^1$ direction) $D0/D4$ system in the limit $j=f_k=0$. Thus it seems that the solution is rightly interpreted (for $f_k=0$) as a $D0/D4$ system in a G\"odel type universe. Given that we showed the existence of supersymmetric solutions describing black holes in G\"odel type universes it is not suprising that supersymmetric D-brane systems in G\"odel type universes exist.

It is interesting to notice that this resolution of causality violations in higher dimensions does not survive to turning on the self-dual part of $fd\omega$: the case $\ell \neq 0$. The ten dimensional metric describing this solution is
\bequ \barr{c} \displaystyle{ds^2=f\left[-dt^2+(dy^1)^2-\left(1-f^{-1}\right)(dt-dy^1)^2+\left\{\left(2g\mp\frac{\ell}{3}r^{\pm2}f^{-1}\right)(dy^1-dt)\right. \right.} \spa{0.3}\\ \displaystyle{\left. \left. \mp\frac{\ell}{3}r^{\pm2}dt\mp \frac{\ell}{6}r^{\pm2}\left(2g\mp\frac{\ell}{6}r^{\pm2}f^{-1}\right)\sigma_{L,R}\right\}\sigma_{L,R}  \right]+f^{-1}ds^2_{\mathbb{R}^4}+ds^2({\mathbb{T}^4})} \earr \ , \eequ
which can still admit non-trivial CTC's since the coefficient of $(\sigma_{L,R})^2$ can be negative.

\subsection{Decoupling limit}
The decoupling limit \cite{Maldacena:1997re} is taken in the string frame, $ds^2=\exp{(\phi/2)}ds^2_E $. First we have to know the quantization of the metric parameters in terms of the string coupling, $g$, the inverse string tension $\alpha'$ and the volume of the compactification $T^4$, $V$, and radius of the circle, $R$. The usual U-duality arguments and quantization of Kaluza-Klein momentum yield
\bequ
Q=\frac{Q_1g(\alpha')^3}{V} \ , \ \ \ \ P=Q_5 g \alpha' \ , \ \  \ \ Q_{KK}=\frac{g^2(\alpha')^4 N_P}{R^2V} \ , \eequ
where $Q_1,Q_5, N_P \in {\mathbb{N}}$. Now we look at the solution (\ref{d15bri}), in the limit where
\bequ
\alpha', r, V\rightarrow 0 \ , \ \ \ \ U\equiv \frac{r}{\alpha'}={\rm{fixed}} \ , \ \ \ \ v\equiv \frac{V}{(\alpha')^2}={\rm{fixed}} \ . \eequ    
In this limit, the resulting solution is
\bequ
\barr{c}
\displaystyle{ds^2_S=\alpha'\left\{\frac{U^2}{\tilde{g}\sqrt{Q_1Q_5}}\left[-dt^2+(dy^1)^2+\frac{\tilde{g}^2N_P}{R^2U^2}(dt-dy^1)^2 +\frac{2j}{(\alpha')^{\pm2}U^{\pm2}}(dy^1-dt)\sigma^3_{L,R}   \right]\right.}\spa{0.4}\\ \displaystyle{\left. + \tilde{g}\sqrt{Q_1Q_5}\left(\frac{dU^2}{U^2}+d\Omega_3^2\right)\right\}} \ ,
\spa{0.6}\\
\displaystyle{C_{RR}=\frac{\alpha' v U^2}{Q_1 g}\left(dt+\frac{j}{(\alpha')^{\pm2}U^{\pm2}}\sigma^3_{L,R}\right)\wedge (dy^1-dt)+\frac{\alpha' g Q_5}{4}\cos{\theta} d\phi_1\wedge d\phi_2 \ , }
\label{d15brideclim}
\earr
\eequ
where $\tilde{g}=g/\sqrt{v}$ and the dilaton is constant. We still have to 
address the $j$ terms in the solution. But before let us do some general considerations on AdS/CFT.

Consider a scalar, vector and tensor perturbation of the  $AdS_3\times S^3$ solution as follows: 
\bequ
ds^2=U^2[-dt^2+(dy^1)^2]+\frac{dU^2}{U^2}+d\Omega_3^2+\frac{T}{U^{n_t}}(dt-dy^1)^2 +\frac{V}{U^{n_v}}(dt-dy^1)d\theta+\frac{S}{U^{n_s}}d\theta^2\ . \label{pertads}\eequ
The mass dimensions of the couplings are 
\bequ
[T]=n_t+2 \ , \ \ \ [V]=n_v+1 \ , \ \ \ [S]=n_s \ .
\eequ
If these couplings act as sources of operators in the dual CFT, we would have the CFT Lagrangian deformed by the terms, respectively,
\bequ
\int d^2x T {\mathcal{O}}_t \ , \ \ \ \int d^2x V {\mathcal{O}}_v \ , \ \ \ \int d^2x S {\mathcal{O}}_s \ . \label{defcft} \eequ
Thus, they would be associated to operators of mass dimension
\bequ
\Delta[{\mathcal{O}}_t]=-n_t \ , \ \ \  \Delta[{\mathcal{O}}_v]=1-n_v \ , \ \ \
\Delta[{\mathcal{O}}_s]=2-n_s \ .
\eequ
But in AdS/CFT, not all perturbations of AdS are associated to deformations of the dual CFT. A given operator in the CFT is associated to two different perturbations in AdS. To see this in more detail consider the unperturbed solution (\ref{pertads}). These coordinates cover a Poincar\'e patch of $AdS_3$. The coordinate transformation $U=1/z$ gives $AdS_3$ in Poincar\'e coordinates $(t,y^1,z)$, where the conformal flatness becomes explicit. The timelike conformal boundary is at $U\rightarrow \infty$, and $U=0$ is a Cauchy horizon. 

If we consider a scalar perturbation of the unperturbed geometry by taking a massive scalar field on $AdS_3$, $\Box \phi=m^2\phi$, it behaves at large U as
\bequ
\phi\sim \frac{a}{U^{1-\sqrt{1+m^2}}}+\frac{b}{U^{1+\sqrt{1+m^2}}} \ . \label{scalarper} \eequ
The first perturbation is non-normalizable (the norm, $\int \sqrt{-g}|\phi|^2$, diverges as $U\rightarrow \infty$); the second one is normalizable.

From the viewpoint of the field theory living on the timelike boundary of $AdS_3$, which is conjectured to be a dual description of the physics in Anti-de-Sitter space \cite{Maldacena:1997re}, $U$ is the energy scale. The two scalar perturbations above correspond to an operator, ${\mathcal{O}}$,  in the CFT. A non-normalizable perturbation in AdS corresponds to deforming the dual field theory, whose Hamiltonian becomes
\bequ
H=H_{CFT}+a{\mathcal{O}} \ , \eequ
whereas a normalizable perturbation corresponds to giving a vev to the operator \bequ
<0|{\mathcal{O}}|0>=b \ . \eequ
Thus we identify the non-normalizable perturbation as being associated to the deformation (\ref{defcft}) and hence we find
\bequ
\Delta[{\mathcal{O}}_s]=1+\sqrt{1+m^2} \ . \eequ
In this scalar case, the operator corresponding to any massive perturbation is an \textit{irrelevant} (in the IR, important in the UV) operator in the CFT since $\Delta>2$. Note that since the CFT is two dimensional marginal operators have dimension $\Delta=2$.

If we consider a massive vector perturbation of the $AdS_3$ geometry, $D_{\mu}F^{\mu \nu}=m^2A^{\nu}$, they behave at large $U$ as (for $m^2\neq 0$) \cite{l'Yi:1998eu,Muck:1998iz}
\bequ A_{U}\sim \frac{c}{U^{3-\sqrt{m^2}}}+\frac{d}{U^{3+\sqrt{m^2}}} \ , \ \ \  \ A_{i}\sim \frac{c_i}{U^{-\sqrt{m^2}}}+\frac{d_i}{U^{+\sqrt{m^2}}} \ , \label{vecper} \eequ where $x^i=(t,y^1)$ \cite{l'Yi:1998eu}. We focus on the $A_i$ perturbations. The first one is non-normalizable. 
The second one is normalizable ($\int \sqrt{-g}g^{ij}A_{i}A_j$ converges on the boundary). Again, identifying the non-normalizable perturbation with the deformation (\ref{defcft}) we find 
\bequ
\Delta[{\mathcal{O}}_v]=1+\sqrt{m^2} \ . \eequ
Thus, a massive, non-normalizable, vector perturbation in $AdS_3$ corresponds to a deformation by a relevant ($0<m^2<1$), marginal ($m^2=1$) or irrelevant ($m^2>1$) operator. The case with $m^2=0$ is more subtle. In Poincar\'{e} coordinates, the solution for vector perturbations in $AdS_3$ is actually
\bequ
A_i\sim K_{\nu}(z)+I_{\nu}(z) \ , \eequ
where these are hyperbolic Bessel functions, and $\nu=\sqrt{m^2}$. For $\nu\neq 0$ these have small $z$ (large U) behavior given by (\ref{vecper}). For $\nu=0$ these behave at small $z$ as
\bequ
A_i\sim -\log{z}+1 \ . \eequ
Thus, a constant behavior is the zero mass limit of the normalizable perturbation. This is the BMPV angular momentum perturbation. Naively, the norm of such constant perturbation seems to diverge logarithmically. However, studying the massless perturbation by introducing a small mass cut-off when computing the norm, and taking the mass to zero at the end, one finds it is indeed a normalizable perturbation. This is in agreement with the known description of the BMPV in terms of CFT states given bellow.

Finally, if we consider a (massless) tensor perturbation. The linearized equation of motion can be found in \cite{Muck:1998ug}. It behaves as we approach the boundary as (in the radiation gauge) \bequ
h_{ij}\sim c_{ij} U^{2}+d_{ij} \ . \label{tenper} \eequ
The first is a non-normalizable mode whereas the second is normalizable. They both correspond to an operator of dimension $\Delta[{\mathcal{O}}_t]=2$, i.e. marginal, which is of course the energy momentum tensor in the CFT.

Our solution in the decoupling limit (\ref{d15brideclim}) has two `perturbation' of $AdS_3\times S^3$. The momentum ($N_P$) `perturbation' is a tensor perturbation, corresponding to the normalizable mode in (\ref{tenper}). It corresponds to giving a vev to the components of the CFT energy momentum tensor $T_{++}$ or $T_{--}$. The angular momentum `perturbations' ($j$) are vector perturbations, which we now discuss in more detail.

\subsubsection{The BMPV case}
For the BMPV case, one quantizes the physical angular momentum of the spacetime (\ref{mjbmpv}). Using $G_5=\pi(\alpha')^4 g^2/(4VR)$, we find
\bequ
j=\frac{N_J(\alpha')^4g^2}{2VR} \  \ \ \Rightarrow \ \ \ \  \frac{j}{(\alpha')^{2}}=\frac{N_J \tilde{g}^2}{2R} \ ,   \eequ
where $N_J \in {\mathbb{N}}$. This angular momentum term is the normalizable mode corresponding to a CFT operator of dimension $\Delta=1$. This is a relevant operator which is the reason why the BMPV entropy is sensitive to the spacetime angular momentum. Moreover, this is a normalizable perturbation, as discussed above. Hence, it corresponds to considering a different set of states (from the $j=0$ case) in the $D1-D5$ CFT \cite{Aharony:1999ti}.

Such states were identified in \cite{BMPV}. The space transverse to the D1-D5 system is four dimensional and has an SO(4) isometry, which translates as the SO(4) R-symmetry of the CFT. Hence the spacetime angular momentum is translated as CFT charge. The two independent angular momentum parameters $\vec{J}=(J_L,J_R)$ translate as the quantized charges, $(F_L,F_R)$, of the $U(1)_L\times U(1)_R$ Cartan sub-algebra of the $SO(4)$ R-symmetry.

The quantum states that describe the static black hole are states with CFT energy, i.e. $(L_0,\tilde{L}_0)$ eigenvalues, $(0,N_P)$. The states that describe the BMPV black hole are `deformed' by carrying charge $(F_L,F_R)=(0,N_J)$ as well. It turns out that the operators that charge the states must carry by unitarity (which in this context means absence of negative norm states) a conformal weight of $3N_J^2/(2c)$, where $c$ is the central charge of the D1-D5 CFT. Thus, by unitarity \cite{Herdeiro:00}
\bequ
N_P>3N_J^2/(2c) \ ,
\eequ
which in terms of the five dimensional black hole means that $\mu^3>4j^2$ and thus we are restricted to the under-rotating case, where all CTC's are hidden behind the event horizon of the black hole.

\subsubsection{The G\"odel black hole case}
The angular momentum term in the decoupling limit metric is, in this case,
\bequ
\alpha'\frac{U^2}{\tilde{g}\sqrt{Q_1Q_5}}\left[2j(\alpha')^{2}U^{2}(dy^1-dt)\sigma^3_{L,R} \right] \ . \eequ 
Since there is no known way to define `physical angular momentum' in G\"odel's universe, we cannot quantize $j$, as we did for the BMPV case. This is because G\"odel-type  solutions are homogeneous spaces and thus, there is no asymptotic region in which the vorticity can be faced as a perturbation, much in the same way that a cosmological constant, however how small, cannot be faced as a perturbation of flat space.  Not quantizing $j$ is consistent with interpreting this term as a deformation of the theory as opposed to an additional quantum number. But this term is sub-leading in $\alpha'$ in the decoupling limit, for finite $j$. Thus, the decoupling limit of the G\"odel black hole solution is the same as the one of the Reissner-Nordstrom black hole, i.e. (\ref{d15brideclim}) with $j=0$. This explains why the black hole entropy is not seeing the spacetime angular momentum at all. Moreover, in the dual CFT, the description of the entropy is exactly the one of the preceding subsection with $N_J=0$.

In order to make the angular momentum term non-vanishing in the decoupling limit, we must take a `double scaling limit' for j
\bequ j\rightarrow \infty \ , \ {\rm{keeping}} \ \ \ j(\alpha')^2\equiv J={\rm{fixed}} \ , \eequ and thus the duality can only say something about G\"odel in the `infinite vorticity limit'. The angular momentum term is the non-normalizable perturbation associated to the insertion of a vector operator  of dimension $\Delta=5$ in the CFT. This operator is irrelevant in the IR, and thus it will not alter the black hole entropy. Notice that such insertion breaks the $SO(4)$ R-symmetry explicitly in the CFT Lagrangian, which is in sharp contrast with the BMPV case.

\section{Conclusions and Discussion}
It has recently been emphasized by Townsend how angular momentum in supersymmetric systems leads to interesting and sometimes surprising configurations \cite{Townsend:2002yf}. In this paper we have shown that a simple, supersymmetric deformation of the D1-D5 system with angular momentum is interpreted in five dimensions as a black hole in a G\"odel-type  universe. The sole existence of a black hole in G\"odel's universe is a novel result and it would be interesting to see if it depends on supersymmetry or not. The entropy of this black hole is `blind' to the spacetime angular momentum and hence this black hole has as many quantum states as the Reissner-Nordstrom black hole of minimal supergravity in five dimensions. This entropy differs, therefore, from the BMPV black hole entropy. We have emphasized all over the text the parallelism between the BMPV and the G\"odel black hole solution; they differ by a sign choice in the two SO(3) algebras in SO(4). 

We have shown that in ten dimensions the `geometric' and non-trivial CTC's of the five dimensional solution become `topological' and trivial. One might therefore ask why should the AdS/CFT duality care about the CTC's if they can be resolved in ten dimensions. Note that the dual CFT description of the five dimensional spacetime requires the wave propagation direction to be compact, in order for the Kaluza-Klein momentum to be quantized, which corresponds to the quantized CFT energy. 

Very recently \cite{Dyson:2002wu} the enhan\c{c}on mechanism was studied in the ten dimensional description of the BMPV black hole and argued that it helps to prevent the appearance of CTC's (in situations where the enhan\c{c}on is relevant, like $K3$ compactifications). From the present paper it is logical to ask what a similar analysis can say about the G\"odel-type universe case. Concerning the `CTC's resolving mechanism' presented herein we would like to add that it does not work for all supersymmetric spinning black holes in five dimensions. A counterexample was given in \cite{Caldarelli:2001iq} for asymptotically AdS black holes. For completeness let us mention that recent work on strings on time dependent backgrounds involves timelike identifications (hence trivial CTC's) in certain parts of flat space (see eg. \cite{Cornalba:2002fi,Liu:2002ft}). After Kaluza-Klein reduction non-trivial CTC's develop, which is exactly the converse of our mechanism. However these are excluded from the lower dimensional manifold since the spacetime develops a curvature singularity before reaching the region with CTC's.

Taking the decoupling limit of the solutions we find two very distinct situations. For the well known BMPV case, the angular momentum term corresponds to a normalizable vector perturbation in $AdS_3$. In the CFT it translates as a dimension $\Delta=1$ operator getting a vacuum expectation value. This operator corresponds to states carrying R-charge, which break spontaneously the R-symmetry. For the G\"odel case, the angular-momentum term is a non-normalizable perturbation. Hence we need not quantize its coefficient since it does not correspond to a quantum number. This term vanishes as we take $\alpha'\rightarrow 0$ unless we take a scaling limit for the spacetime vorticity. This explains why the CFT will be `blind' to the finite spacetime vorticity. It would be interesting to understand if the infinite vorticity deformation of the CFT can be ruled out by unitarity.

Finally, to understand the properties of the solutions (\ref{solR4}) with $\ell\neq 0$ both in 5 and 10 dimensions is an open question.

\section *{Acknowledgments}
I would like to thank R.Emparan, J.Lemos and C.Nu\~nez for discussions and G.Gibbons and P.Townsend for correspondence. I am especially grateful to M.Costa, S.Hirano and H.Reall for many suggestions. The author is supported by the grant SFRH/BPD/5544/2001 (Portugal).

\end{document}